\documentclass[pra,showpacs,preprintnumbers,amsmath,aps]{revtex4}
\usepackage{graphics}

\def\be{\begin{equation}}
\def\ee{\end{equation}}
\def\bea{\begin{eqnarray}}
\def\eea{\end{eqnarray}}
\def\bse{\begin{subequations}}
\def\ese{\end{subequations}}
\def\bma{\begin{mathletters}}
\def\ema{\end{mathletters}}
\def\C{\hbox{$\mit I$\kern-.6em$\mit C$}}

\begin{document}

\title{Distinguishing maximally entangled states locally}
\author{Dong Yang}
\email{dyang@zju.edu.cn} 
\author{Yi-Xin Chen}
\affiliation{Zhejiang Institute of Modern Physics and
Department of Physics, Zhejiang University, Hangzhou 310027, P. R. China}

\date{\today}

\begin{abstract}
We demonstrate that one maximally entangled state is sufficient and necessary to distinguish a complete basis of maximally entangled states by local operation and classical communication.  
\end{abstract}


\maketitle

Generally orthogonal states may be distinguished perfectly only by means of global measurements since quantum information of orthogonality may be encoded in entanglement that cannot be extracted by local operation and classical communication (LOCC). Whether or not a set of orthogonal states can be discriminated by LOCC is related closely to the number of the states. Any two orthogonal multipartite states could be discriminated with certainty by only LOCC operations \cite{Walgate1} while there exist a complete basis of product orthogonal states that could not \cite{Bennett}. An arbitrary complete set of orthogonal states of any bipartite system is locally indistinguishable if at least one of the vectors is entangled \cite{Horodecki}. To locally discriminate the complete basis of maximally entangled states, entanglement is required. In this note, We prove that a maximally entangled state is sufficient and necessary.

We consider the problem of distinguishing the canonical set of mutually orthogonal maximally entangled states (MES) in $d\times d$ defined as,
\be
|\phi_{m,n}\rangle=\frac{1}{\sqrt{d}}\sum_{j=0}^{d-1}e^{2\pi ijn/d}|j\rangle|j\oplus m\rangle, n,m=0,1,\cdots,d-1,
\ee
where $\oplus$ means addition modulo $d$. These orthogonal states cannot be distinguished exactly by LOCC operations. To distinguish them, entanglement is required.  

{\bf Main Result}: The canonical set of mutually orthogonal maximally entangled states (MES) in $d\times d$ is distinguished locally iff one pure MES is shared by the two parties.
 
Proof:
The sufficient condition is easy to prove. Employing the standard teleportation protocol \cite{Bennett2}, the state is teleported to one party by the shared pure MES and then general Bell measurement is performed to distinguish the state perfectly. The necessary condition is proved using the entanglement theory that entanglement cannot increased by LOCC operations. Similar method is used to investigate the problem of local cloning of Bell states in \cite{Ghosh}. Suppose a entangled state $\psi$ is required to distinguish the MES. Then a mixed state is construct as follows
\be
\rho=\frac{1}{d^{2}}\sum_{m,n=0}^{d-1}|\phi_{m,n}\rangle\langle\phi_{m,n}|\otimes|\psi\rangle\langle\psi|\otimes|\phi_{m,-n}\rangle\langle\phi_{m,-n}|.
\ee 
The distillable entanglement of $\rho$ is at least $\log d$ since we can distinguished $|\phi_{m,n}\rangle|\psi\rangle$ by LOCC. Note that the state $\rho_{s}$     
\be
\rho_{s}=\frac{1}{d^{2}}\sum_{m,n=0}^{d-1}|\phi_{m,n}\rangle\langle\phi_{m,n}|\otimes|\phi_{m,-n}\rangle\langle\phi_{m,-n}|,
\ee
is separable \cite{Smolin,Fan}. So the distillable entanglement of $\rho$ is equal to the entanglement of $\psi$ that is at least $\log d$. From the sufficient condition, $\psi$ is a MES in $d\times d$. The proof is completed.

D. Yang thanks S. J. Gu for helpful discussion. 


\end{document}